\documentclass[11pt]{article}
\usepackage[version=4]{mhchem}
\usepackage{xr}
\usepackage{graphicx}
\graphicspath{ {./figures/} }
\usepackage[section]{placeins}
\usepackage{float}
\usepackage[super,comma,sort&compress]{natbib}
\usepackage[singlelinecheck=false]{caption}
\floatstyle{plaintop}
\restylefloat{table}
\usepackage{multicol}
\usepackage{geometry}
\usepackage{physics}
\usepackage{xcolor}
\usepackage[implicit=false, hidelinks]{hyperref}
\usepackage{amsmath}
\usepackage{amssymb}
\usepackage{indentfirst}
\usepackage{authblk}

\makeatletter
\newcommand*{\addFileDependency}[1]{
  \typeout{(#1)}
  \@addtofilelist{#1}
  \IfFileExists{#1}{}{\typeout{No file #1.}}
}
\makeatother

\usepackage{lineno}

\begin{document}

\title{AdsNet: A Graph Neural Network for Direct Adsorption Energy Prediction}
\author[1,2]{Changzhi Ai\thanks{Corresponding author: changzhi@stanford.edu}}
\author[2]{Frank Abild-Pedersen}
\author[2]{Johannes Voss}


\affil[1]{Department of Chemical Engineering, Stanford University, Stanford, California 94305, United States}
\affil[2]{SUNCAT Center for Interface Science and Catalysis, SLAC National Accelerator Laboratory, Menlo Park, California 94025, United States}
\date{}

\maketitle

\begin{abstract}
Fast and accurate prediction of adsorption energies is critical for high-throughput computational catalyst screening, existing methods compromise either on evaluation speed (full-graph machine learning potentials) or geometric sensitivity (coordination-based models like the $\alpha$-scheme). Here, we present AdsNet, an equivariant graph neural network that directly predicts adsorption energies in a single forward pass, which operates on a local graph restricted only to adsorbate atoms and its immediate surface neighbors within a cutoff radius. This local graph formulation decouples computational complexity from total system size while encoding continuous geometric features (interatomic distances and directional vectors) enabling sensitivity to subtle variations in local bonding environments. AdsNet adopts a message-passing architecture with scalar and vector node features over multiple interaction layers. Trained on AdsTraj, a dataset of 33{,}920 density functional theory (DFT) calculated structures spanning 26 transition metals and 6 adsorbates (H*, C*, O*, CO*, OH*, HOCO*), the model achieves an adsorption energy root-mean-square error (RMSE) of 0.032~eV and a force RMSE of 0.047~eV/\AA{}. Analytic forces obtained via automatic differentiation enable structure relaxation through a projected gradient descent algorithm with adaptive line search, eliminating entire geometry optimizations. Additionally, AdsNet uniquely decomposes predicted adsorption energies into contributions from individual atoms within the local adsorbate graph—an interpretability feature unattainable with conventional machine-learning potentials. AdsNet thus establishes an accurate, highly efficient, and interpretable framework for large-scale adsorption energy prediction and local interface relaxation across various material surfaces.

\end{abstract}

\section{Introduction}

Heterogeneous catalysis plays an essential role in energy conversion and chemical synthesis.\cite{Norskov2014,Friend2017} It underpins a wide range of industrial processes responsible for the production of fuels, fertilizers, and commodity chemicals, while playing a critical role in global energy efficiency and environmental sustainability. As the global transition toward a low-carbon and sustainable energy economy accelerates, the development of advanced heterogeneous catalysts has become increasingly important for emerging electrochemical technologies, including water splitting for hydrogen production, carbon dioxide reduction to value-added fuels and chemicals, and nitrogen fixation for ammonia synthesis.\cite{Nitopi2019,Roger2017,Stamenkovic2016,Ai2022,Singh2016} These technologies offer promising pathways for renewable energy storage and utilization, but their widespread implementation is often limited by low reaction activity, poor selectivity, and catalyst degradation under operating conditions. The performance of a heterogeneous catalyst is fundamentally governed by the interactions between reaction intermediates and active surface sites. Small variations in surface composition, atomic arrangement, coordination environment, or electronic structure can significantly influence reaction energetics, catalytic activity, and product selectivity.\cite{Nrskov2008,Medford2015,Ai2023,Ai2024} Consequently, understanding how reactants, intermediates, and products adsorb and react on catalyst surfaces at the atomic scale is essential for elucidating reaction mechanisms and identifying the factors that control catalytic performance. These atomic insights establish the robust structure–property relationships that guide rational catalyst design.

The adsorption energies of reactants, intermediates, and products on catalyst surfaces are among the most important descriptors governing catalytic activity, selectivity, and stability.\cite{Norskov2011,Nitopi2019} According to the Sabatier principle, optimal catalysts bind key reaction intermediates neither too strongly nor too weakly, resulting in maximum catalytic performance.\cite{Norskov2011} This concept underlies the design of catalysts for a wide range of electrochemical reactions, including the carbon dioxide reduction reaction (CO$_2$RR), hydrogen evolution reaction (HER), oxygen evolution reaction (OER), nitrogen reduction reaction (NRR), and so on. Consequently, accurate determination of adsorption energies is critical for understanding reaction mechanisms, establishing activity descriptors, and guiding the rational design of catalytic materials. Computational catalyst screening based on density functional theory (DFT) has emerged as a powerful tool for accelerating catalyst discovery by predicting adsorption energetics and catalytic properties from first principles.\cite{Frank2007,Jaramillo2007,Norskov2011,Andersen2021} Over the past two decades, DFT calculations have enabled the development of adsorption-energy scaling relations, activity volcano plots, and descriptor-based design frameworks that have significantly advanced the understanding of heterogeneous catalysis. Compared with experimental trial-and-error approaches, DFT provides atomic-scale insight into catalyst--adsorbate interactions and enables systematic exploration of catalyst composition, surface structure, and reaction pathways.

Despite its success, DFT remains computationally demanding, particularly for large-scale screening studies involving thousands of catalyst surfaces, adsorbates, and reaction intermediates. The high computational cost of geometry optimization and adsorption-energy calculations limits the exploration of the vast compositional and configurational space of catalytic materials. As a result, developing accurate and computationally efficient methods for predicting adsorption energies has become a critical challenge in computational catalysis. Recent advances in machine learning (ML) have provided promising opportunities to overcome these limitations. In particular, machine learning interatomic potentials (MLIPs) have emerged as powerful tools for accelerating atomistic simulations while retaining near-DFT accuracy.\cite{Schlexer2019,Behler2007,Deringer2021} Among the various MLIP architectures, equivariant message-passing neural networks (MPNNs) have demonstrated remarkable performance due to their ability to explicitly incorporate the rotational, translational, and permutation symmetries of atomic systems.\cite{Deringer2021} By representing atomic environments, these models can accurately capture complex many-body interactions in molecules and materials. Representative examples include PaiNN,\cite{Kristof2021} NequIP\cite{Batzner2022}, Allegro\cite{Musaelian2023}, MACE\cite{Batatia2022}, Equiformer\cite{liao2022}, and UMA\cite{UMA2025}, which have achieved state-of-the-art accuracy across a wide range of atomistic prediction tasks. All of these MLIPs are designed to predict total energies and atomic forces for periodic and isolated atomic structures. Their application to adsorption energetics relies on calculating the energy difference between the adsorbate–surface complex, the clean surface and gas-phase molecule. This scheme inherently requires two separate model evaluations -- one for the bare surface and another for the combined adsorbate–surface system. Alpha scheme ($\alpha$-scheme) is another type of physics-based model that estimates adsorption energies based on the coordination numbers of adsorbate on metal surfaces,\cite{Roling2017,Roling2018,Saini2024} which has been successfully applied in metal systems. The $\alpha$-scheme estimates adsorption energies across a broad range of adsorbates on metal surfaces by solving a system of linear equations parameterized from reference DFT data. Owing to its compact parameter set, the model is computationally instantaneous, highly interpretable, and directly yields adsorption energies in a single step without requiring separate evaluations of the clean slab and the adsorbed complex.

Although the aforementioned equivariant MLIPs achieve remarkable accuracy in predicting total energies and atomic forces, adapting them to adsorption-energy prediction presents several practical challenges. These models typically construct graphs containing all atoms in the system, including all non-neighbor atoms, resulting in computational costs that scale with the total number of atoms and edges. For typical slab models containing tens to hundreds of surface atoms, the resulting graphs may contain thousands to tens of thousands of edges, leading to substantial computational overhead. Furthermore, adsorption energies are generally obtained as energy differences between the adsorbate–surface complex and the corresponding clean surface, requiring separate model evaluations for each structure. Consequently, the computational cost of adsorption-energy prediction is effectively doubled, which can become a significant bottleneck in large-size particles and large-scale catalyst screening studies involving thousands or millions of candidate adsorption configurations. In contrast, the $\alpha$-scheme directly predicts adsorption energies in a single evaluation with negligible computational cost, and then enable rapid screening of catalyst surfaces based on local coordination environments. However, the simplicity of the $\alpha$-scheme also imposes significant limitations. Because adsorption energies are parameterized solely through discrete coordination descriptors, the $\alpha$-scheme model is insensitive to continuous geometric variations such as bond lengths, bond angles, adsorbate orientations and surface relaxation effects. As a result, adsorption configurations with identical coordination numbers but distinct local geometries are assigned the same predicted energy. In addition, the $\alpha$-scheme does not provide atomic force information, preventing its application to geometry optimization or structural relaxation. These limitations severely constrain its ability to capture the nuanced geometric effects that govern adsorption energetics on realistic catalyst surfaces.

In this work, we present AdsNet, an equivariant graph neural network designed to address these limitations by uniting the advantages of equivariant MLIPs and the $\alpha$-scheme for direct adsorption-energy prediction. AdsNet directly predicts adsorption energies in a single forward pass without requiring separate evaluations for the bare surface and the adsorbate–surface complex. By constructing a localized graph restricted strictly to the adsorbate atoms and their immediate surface neighbors within a defined cutoff radius, AdsNet decouples its computational complexity from the overall system size. This local graph encodes continuous geometric features---interatomic distances and directional vectors---enabling high sensitivity to subtle variations in local bonding environments. Furthermore, AdsNet natively decomposes the predicted adsorption energy into contributions from individual atoms within the local adsorbate graph, unlocking atomic-scale interpretability that MLIPs cannot provide. Leveraging automatic differentiation, AdsNet can also yield atomic forces for direct local structure relaxation. Here, we detail the model architecture, training protocol, and validation on DFT relaxations, showing that AdsNet delivers near-DFT accuracy, interpretable energetics, and DFT-free structural relaxation with minimal computational overhead compared to full-graph MLIPs.

\section{Related Work}

As introduced above, the $\alpha$-scheme is a physics-based model that estimates adsorption energies directly from the coordination numbers of the adsorbate on a metal surface.\cite{Roling2017,Roling2018,Saini2024} Here we describe its formulation in more detail. The scheme expresses the adsorption energy of an adsorbate as a linear combination of its coordination numbers, which gives a system of linear equations:
\begin{equation}
    A \vec{\alpha} = \vec{E_b}
\end{equation}
where $A$ is the coordination number matrix, $\vec{E_b}$ is the vector of adsorption energies, and $\vec{\alpha}$ is the vector of $\alpha$ parameters, obtained by solving this system against reference DFT calculations. This compact formulation has been applied successfully across a wide range of adsorbates and metal surfaces, making it a valuable tool for catalyst screening and design. A key advantage of the $\alpha$-scheme is that its physics-based form is more interpretable, and potentially more generalizable, than purely data-driven models. However, deriving the necessary parameters may require manually selecting DFT calculations spanning different coordination-number configurations. The parameter set is also small, typically 12--14 for each metal, which may limit its accuracy for complex systems. There are no distance- or angle-dependent parameters in the $\alpha$-scheme, which makes it insensitive to the distance between the adsorbate and the surface and to the orientation of the adsorbate. In addition, the $\alpha$-scheme does not provide force predictions, which limits its applicability for structure relaxation.

MLIPs, by contrast, learn total energies and atomic forces directly from DFT reference data at a small fraction of the computational cost. 
High-dimensional neural network potentials (HDNNPs) and atom-centered symmetry functions are among the earliest MLIPs that have been successfully applied to a wide range of systems.\cite{Behler2007,Behler2011} While these models can achieve good accuracy, they often require a large number of parameters and do not explicitly incorporate physical equivariant symmetries. They usually use feed-forward neural networks or Gaussian processes with atomic numbers, distances, and angles invariant features, which can limit their accuracy and generalizability. Message-passing neural networks\cite{Gilmer2017,Yang2024} address this limitation by building the symmetries of atomic systems directly into the network, so that predicted energies are invariant under rotation and translation while predicted forces transform equivariantly. This allows them to capture the directional nature of bonding in adsorbate–surface systems more effectively than invariant descriptors alone. These equivariant architectures have demonstrated state-of-the-art performance in predicting energies and forces for molecular and materials systems. Based on MLIPs, the adsorption energy is obtained as the energy difference between the adsorbate–surface system, the clean surface, and the gas-phase adsorbate:
\begin{equation}
    E_\mathrm{ads} = E_\mathrm{surf+ads} - E_\mathrm{surf} - E_\mathrm{ref}
\end{equation}
where $E_\mathrm{surf+ads}$ is the energy of the adsorbate–surface system, $E_\mathrm{surf}$ is the energy of the clean surface, and $E_\mathrm{ref}$ is the energy of the gas-phase adsorbate. By using MLIPs to predict the energies of these three systems, we can efficiently calculate the adsorption energy with high accuracy, which can be several orders of magnitude faster than DFT calculations. 

AdsorbML has been proposed as an efficient workflow for adsorption energy calculations by combining generalizable machine learning potentials with selective DFT relaxations.\cite{AdsorbML2023} By using a pre-trained GNN to generate candidate adsorbate configurations and relaxing only the most promising ones with DFT, AdsorbML achieves a 2--3 order of magnitude speedup over conventional DFT-only approaches while retaining high accuracy across a wide range of adsorbate–surface systems. Compared to adsorption energy prediction with MLIPs, AdsorbML focuses more on the workflow and strategy for accurately and efficiently calculating adsorption energies. It is a promising method to trade off computational efficiency for accuracy. However, AdsorbML still relies on DFT calculations for structure relaxation, which can be a bottleneck for large-scale screening.

\section{Results and Discussion}

\subsection{Adsorbate Graph Representation}

\FloatBarrier
\begin{figure}[!ht]
    \centering
    \includegraphics[width=0.5\textwidth]{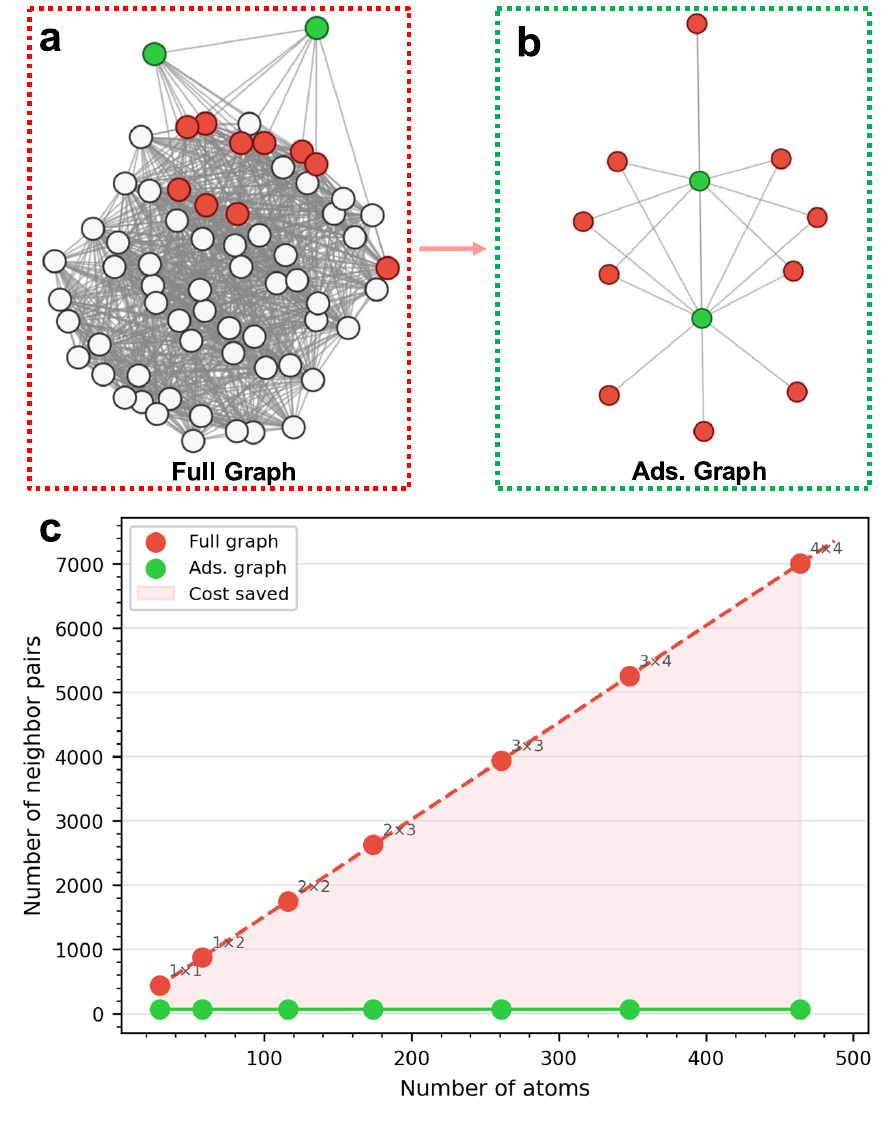}
    \caption{Transforming (a) full graph to (b) adsorbate graph. (c) Computational cost comparison between full graph and adsorbate graph.}
    \label{fig:cost}
\end{figure}

DFT calculations for adsorption energies typically require several hours to days of computational time per structure, depending on the system size and computational resources. MLIPs that learn from DFT data have been shown to significantly reduce computational cost compared to DFT, and keep accurate predictions of energy, forces, and so on. Especially, Graph Neural Networks (GNNs) equipped with equivariant message passing have demonstrated promising results in this regard, like PaiNN, NequIP, MACE, Allegro, UMA and so on. All of these architectures first generate full graphs of the geometric structure to represent the system. Figure \ref{fig:cost}a shows a full graph representation of Pt $4\times4\times4$ with one CO* adsorbate. Each node in the graph represents an atom, and each edge represents a bond between atoms within a cutoff radius. There are 66 nodes, and 1083 edges in this full graph. Among these nodes, the green node represents the adsorbate, and the rest of the nodes represent the surface atoms. The full graph representation captures all the interactions between atoms in the system, which is crucial for accurate energy and force predictions. However, it also produces a large number of edges, which keeps computational costs high during both training and inference. In contrast, Figure \ref{fig:cost}b shows the localized adsorbate graph (Ads. Graph) representation adopted in AdsNet. This local adsorbate graph only includes nodes and edges between the adsorbate and the surface atoms. For the same Pt surface, the adsorbate graph consists of 2 adsorbate nodes (green), 10 surface nodes (red), and 17 edges, which are only a very small part of the full graph. This significantly reduces the number of nodes from 66 to 12, and the number of edges from 1083 to 17, while still capturing the most important interactions for adsorption energy prediction.  By using this simplified graph representation, AdsNet can achieve much faster training and inference compared to full graph-based GNNs, which can be intuitively seen from the computational cost comparison in Figure \ref{fig:cost}c. The number of edges in the full graph scales linearly with the number of atoms, while the number of edges in the adsorbate graph is independent of the number of surface atoms. This leads to a dramatic reduction in computational cost shifting from full graph to adsorbate graph, especially for larger systems.

\subsection{Model Architecture}

\FloatBarrier
\floatstyle{plain}
\restylefloat{figure}
\begin{figure}[!ht]
    \centering
    \includegraphics[width=0.6\textwidth]{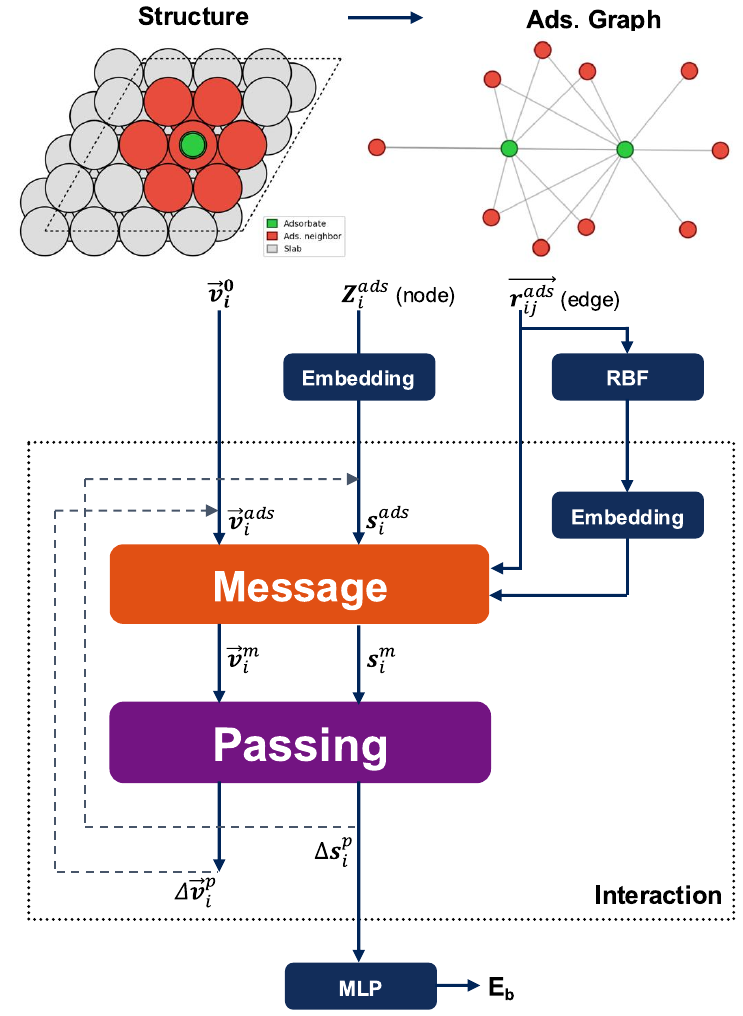}
    \caption{Architecture of AdsNet. The model uses equivariant message passing with scalar and vector node features over $L$ interaction layers, followed by a readout MLP to predict adsorption energies.}
    \label{fig:arch}
\end{figure}

AdsNet implements an equivariant message-passing architecture that, unlike conventional GNN-based potentials, operates on a localized adsorbate graph rather than a full-system graph as the input representation. Given an input structure as illustrated in Figure \ref{fig:arch}, only the local adsorbate atoms and surface atoms within a cutoff radius (highlighted in green and red) are represented as a graph (adsorbate graph), rather than all atoms in the system. This localized adsorbate graph contains exclusively the adsorbate nodes (in green), their neighboring surface nodes (in red), and the edges connecting them. The inputs of the architecture are atomic numbers $\mathbf{Z}_i^{ads}$ within the adsorbate graph and the interatomic vectors $\overrightarrow{\mathbf{r}}_{ij}^{ads}$ between the adsorbate and surface atoms. After the embedding of atomic numbers, it is projected to scalar features. $\overrightarrow{\mathbf{r}}_{ij}^{ads}$ on the one hand is directly passed to message block to contribute to vector features by a scalar interaction. On the other hand, its interatomic distances are expanded using radial basis functions (RBF) and then through embedding to contribute to scalar features by a cosine cutoff envelope. AdsNet supports two formulations for the radial basis functions: the Gaussian RBF and the sinc RBF. The Gaussian RBF is defined as:
\begin{equation}
    \phi_n(r) = \exp\left(-\frac{(r - \mu_n)^2}{2\sigma^2}\right), \quad n = 1, \ldots, N_\mathrm{basis}
    \label{eqn:gaussian_rbf}
\end{equation}
where $\mu_n$ are the centers of the Gaussian functions and $\sigma$ is the width parameter. The sinc RBF can be defined as:
\begin{equation}
    \phi_n(r) = \frac{\sin(n \pi r / r_\mathrm{cut})}{r}, \quad n = 1, \ldots, N_\mathrm{basis}
    \label{eqn:sinc_rbf}
\end{equation}
where $r_\mathrm{cut}$ is the cutoff distance. 
The choice between these radial basis functions directly influences model performance and can be treated as a tunable hyperparameter during training. The cosine cutoff envelope is defined as:
\begin{equation}
    f_\mathrm{cut}(r) = \frac{1}{2}\left(\cos\left(\frac{\pi r}{r_\mathrm{cut}}\right) + 1\right), \quad r < r_\mathrm{cut}
    \label{eqn:cutoff}
\end{equation}

Each atom $i$ within the adsorbate graph is represented by a scalar feature $\mathbf{s}_i^{ads} \in \mathbb{R}^{F}$ and a vector feature $\mathbf{v}_i^{ads} \in \mathbb{R}^{3 \times F}$ (initialized as zero vector $\overrightarrow{\mathbf{v}}_{ij}^{0}$), where $F$ is the number of channels. Then the scalar and vector features are updated through $L$ interaction layers. Each interaction layer consists of a message block and an update block, which are referenced from the PaiNN architecture\cite{Kristof2021}. The message block computes messages from neighboring nodes and edges, while the update block updates the node features based on the messages.

\textbf{Message block.} For each edge $(j \to i)$, the filter weight is computed from the radial basis embedding modulated by a cosine cutoff envelope:
\begin{equation}
    \mathbf{W}_{ij}^{ads} = \mathbf{W}_\phi(\mathbf{e}_{ij}^{ads}) \cdot f_\mathrm{cut}(r_{ij}^{ads})
    \label{eqn:filter}
\end{equation}
where $\mathbf{W}_\phi$ is the edge embedding network, $\mathbf{e}_{ij}^{ads}$ is the radial basis expansion of the interatomic distance $r_{ij}^{ads}$, and $f_\mathrm{cut}$ is the cosine cutoff function. The filter weight is split into three components and applied to a scalar MLP to produce a gated scalar message, a gated vector from the neighbor's state, and a gated edge direction vector:
\begin{equation}
    \Delta \mathbf{s}_i^{ads} = \sum_{j \in \mathcal{N}(i)} \mathbf{W}_{ij}^{ads,(s)} \odot \mathrm{MLP}(\mathbf{s}_j^{ads})
    \label{eqn:message_s}
\end{equation}
\begin{equation}
    \Delta \vec{\mathbf{v}}_i^{ads} = \sum_{j \in \mathcal{N}(i)} \left[ \mathbf{W}_{ij}^{ads,(v)} \odot \vec{\mathbf{v}}_j^{ads} + \mathbf{W}_{ij}^{ads,(e)} \odot \hat{\mathbf{r}}_{ij}^{ads} \right]
    \label{eqn:message_v}
\end{equation}
where $\hat{\mathbf{r}}_{ij}^{ads} = \overrightarrow{\mathbf{r}}_{ij}^{ads} / r_{ij}^{ads}$ is the unit direction vector. The three gate components $\mathbf{W}_{ij}^{ads,(s)}$, $\mathbf{W}_{ij}^{ads,(v)}$, and $\mathbf{W}_{ij}^{ads,(e)}$ are obtained by splitting the element-wise product of the filter weight and the scalar MLP output into three equal parts.

\textbf{Update block.} The scalar and vector features are updated via an equivariant update following the PaiNN architecture\cite{Kristof2021}. Two linear projections $\mathbf{U}$ and $\mathbf{V}$ are applied to the vector features, and a shared MLP takes the concatenation of $\|\mathbf{V}\vec{\mathbf{v}}_i^{ads}\|$ and $\mathbf{s}_i^{ads}$ to produce three scalar gates $a_{vv}$, $a_{sv}$, and $a_{ss}$:
\begin{equation}
    \mathbf{s}_i^{ads} \leftarrow \mathbf{s}_i^{ads} + a_{ss} + a_{sv} \langle \mathbf{U}\vec{\mathbf{v}}_i^{ads},\, \mathbf{V}\vec{\mathbf{v}}_i^{ads} \rangle
    \label{eqn:update_s}
\end{equation}
\begin{equation}
    \vec{\mathbf{v}}_i^{ads} \leftarrow \vec{\mathbf{v}}_i^{ads} + a_{vv}\, \mathbf{U}\vec{\mathbf{v}}_i^{ads}
    \label{eqn:update_v}
\end{equation}

\textbf{Readout.} Adsorption energy contributions are predicted by a two-layer MLP applied to the final scalar features, and the total adsorption energy is obtained by summing over atoms within adsorbate graph:
\begin{equation}
    E_\mathrm{ads} = \sum_{i \in \mathrm{ads}} \mathrm{MLP}(\mathbf{s}_i^{ads,(L)})
    \label{eqn:readout}
\end{equation}

\textbf{Forces.} Atomic forces are obtained as the negative gradient of the predicted adsorption energy with respect to atomic positions:
\begin{equation}
    \mathbf{F}_i = -\frac{\partial E_\mathrm{ads}}{\partial \mathbf{r}_i}
    \label{eqn:forces}
\end{equation}
Because AdsNet parametrizes $E_\mathrm{ads}$ as a differentiable function of atomic coordinates defined on the localized adsorbate graph, these atomic forces are evaluated analytically via automatic differentiation. These values reflect the force differential relative to the fully relaxed equilibrium geometry, where residual net forces vanish.

\subsection{Dataset}

\begin{table}[!ht]
\centering
\caption{Number of structures per metal and adsorbate in the dataset. Crystal structure and exposed facet are listed for each metal.}
\label{tab:dataset}
\small
\renewcommand{\arraystretch}{0.7}
\begin{tabular}{l c c r r r r r r r}
\hline
Metal & Structure & Facet & H* & C* & O* & CO* & OH* & HOCO* & Total \\
\hline
Sc  & hcp & (0001) & 84  & 164 & 143 & 225 & 213 & 663  & 1492 \\
Ti  & hcp & (0001) & 78  & 83  & 51  & 313 & 237 & 469  & 1231 \\
V   & bcc & (110)  & 51  & 76  & 92  & 212 & 169 & 528  & 1128 \\
Cr  & bcc & (110)  & 43  & 67  & 68  & 224 & 182 & 526  & 1110 \\
Mn  & fcc & (111)  & 99  & 64  & 96  & 229 & 221 & 576  & 1285 \\
Fe  & bcc & (110)  & 96  & 110 & 48  & 244 & 194 & 672  & 1364 \\
Co  & hcp & (0001) & 75  & 78  & 91  & 332 & 208 & 459  & 1243 \\
Ni  & fcc & (111)  & 84  & 100 & 77  & 163 & 186 & 482  & 1092 \\
Cu  & fcc & (111)  & 82  & 95  & 118 & 228 & 276 & 682  & 1481 \\
Zn  & hcp & (0001) & 101 & 138 & 139 & 689 & 338 & 677  & 2082 \\
Y   & hcp & (0001) & 72  & 113 & 132 & 406 & 236 & 657  & 1616 \\
Zr  & hcp & (0001) & 78  & 92  & 74  & 344 & 180 & 572  & 1340 \\
Nb  & bcc & (110)  & 41  & 110 & 106 & 212 & 89  & 756  & 1314 \\
Mo  & bcc & (110)  & 51  & 72  & 63  & 178 & 105 & 659  & 1128 \\
Ru  & hcp & (0001) & 74  & 73  & 73  & 97  & 153 & 406  & 876  \\
Rh  & fcc & (111)  & 80  & 95  & 79  & 306 & 154 & 478  & 1192 \\
Pd  & fcc & (111)  & 106 & 91  & 55  & 172 & 174 & 429  & 1027 \\
Ag  & fcc & (111)  & 95  & 86  & 95  & 168 & 247 & 924  & 1615 \\
Cd  & hcp & (0001) & 122 & 202 & 216 & 686 & 442 & 839  & 2507 \\
Hf  & hcp & (0001) & 82  & 80  & 73  & 245 & 153 & 464  & 1097 \\
Ta  & bcc & (110)  & 55  & 88  & 80  & 255 & 80  & 542  & 1100 \\
W   & bcc & (110)  & 49  & 69  & 62  & 289 & 89  & 491  & 1049 \\
Re  & hcp & (0001) & 72  & 117 & 122 & 204 & 194 & 411  & 1120 \\
Ir  & fcc & (111)  & 47  & 90  & 72  & 162 & 118 & 419  & 908  \\
Pt  & fcc & (111)  & 47  & 116 & 56  & 176 & 127 & 513  & 1035 \\
Au  & fcc & (111)  & 120 & 158 & 126 & 285 & 196 & 603  & 1488 \\
\hline
Total &   &        & 1984 & 2627 & 2407 & 7044 & 4961 & 14897 & 33920 \\
\hline
\end{tabular}
\renewcommand{\arraystretch}{0.9}
\end{table}

To train and benchmark AdsNet, we developed AdsTraj,\cite{AdsTraj2025} a comprehensive dataset of DFT-calculated adsorption energies encompassing various adsorbates on various transition-metal surfaces across the 3$d$ (Sc--Zn), 4$d$ (Y--Cd), and 5$d$ (Hf--Au) series. The six adsorbates considered are H*, C*, O*, CO*, OH*, and HOCO*, which are key intermediates across key electrochemical reactions including HER, CO$_2$RR, and OER. Surface slabs are constructed based on experimental ground-state crystal structures of each metal: face-centered cubic (fcc) metals are modeled with the (111) facet, body-centered cubic (bcc) metals with the (110) facet, and hexagonal close-packed (hcp) metals with the (0001) facet. Each slab is modeled using a $3 \times 3$ surface supercell with three atomic layers and a 10~\AA{} vacuum region. The bottom two atomic layers are constrained to their bulk coordinates, whereas the uppermost layer and the adsorbate are fully relaxed. Multiple adsorption sites are considered for each metal--adsorbate combination, encompassing ontop, bridge, fcc-hollow, and hcp-hollow sites for fcc and hcp facets, alongside ontop, long-bridge, short-bridge, and hollow sites for bcc facets.

All DFT calculations are performed using the Vienna Ab initio Simulation Package (VASP) within the RPBE exchange-correlation framework.\cite{Kresse1993,Kresse1994,Kresse1996,Hammer1999} Electronic wavefunctions are expanded with a plane-wave kinetic energy cutoff of 400~eV, and reciprocal space is sampled on a $3 \times 3 \times 1$ Monkhorst-Pack $k$-point grid. Ionic relaxations are converged until interatomic forces fall below 0.01~eV/\AA{} alongside an electronic energy convergence threshold of $10^{-5}$~eV. In total, the resulting AdsTraj dataset encompasses 33{,}920 structures extracted from these relaxation trajectories. Adsorption energies are computed as $E_\mathrm{ads} = E_\mathrm{slab+ads} - E_\mathrm{slab} - E_\mathrm{ref}$, where $E_\mathrm{ref}$ is the gas-phase reference energy of each adsorbate. The reference energies for the four gas-phase molecules are obtained from separate DFT calculations:
\begin{align}
    E_\mathrm{ref}(\ce{CO2}) &= -22.273~\mathrm{eV}, \quad
    E_\mathrm{ref}(\ce{H2}) = -6.985~\mathrm{eV}, \nonumber \\
    E_\mathrm{ref}(\ce{O2}) &= -8.466~\mathrm{eV}, \quad
    E_\mathrm{ref}(\ce{H2O}) = -14.147~\mathrm{eV}
    \label{eqn:gas_refs}
\end{align}
From these calculations, the reference energy for each adsorbate species is derived according to the following equations:
\begin{align}
    E_\mathrm{ref}(\ce{H^*}) &= \tfrac{1}{2}\, E_\mathrm{ref}(\ce{H2}) \label{eqn:ref_H} \\
    E_\mathrm{ref}(\ce{O^*}) &= E_\mathrm{ref}(\ce{H2O}) - E_\mathrm{ref}(\ce{H2}) \label{eqn:ref_O} \\
    E_\mathrm{ref}(\ce{C^*}) &= E_\mathrm{ref}(\ce{CO}) - E_\mathrm{ref}(\ce{H2O}) + E_\mathrm{ref}(\ce{H2}) \label{eqn:ref_C} \\
    E_\mathrm{ref}(\ce{CO^*}) &= E_\mathrm{ref}(\ce{CO}) \label{eqn:ref_CO} \\
    E_\mathrm{ref}(\ce{OH^*}) &= E_\mathrm{ref}(\ce{H2O}) - \tfrac{1}{2}\, E_\mathrm{ref}(\ce{H2}) \label{eqn:ref_OH} \\
    E_\mathrm{ref}(\ce{HOCO^*}) &= E_\mathrm{ref}(\ce{CO2}) + \tfrac{1}{2}\, E_\mathrm{ref}(\ce{H2}) \label{eqn:ref_HOCO}
\end{align}

\FloatBarrier
\begin{figure}[!ht]
    \centering
    \includegraphics[width=1.0\textwidth]{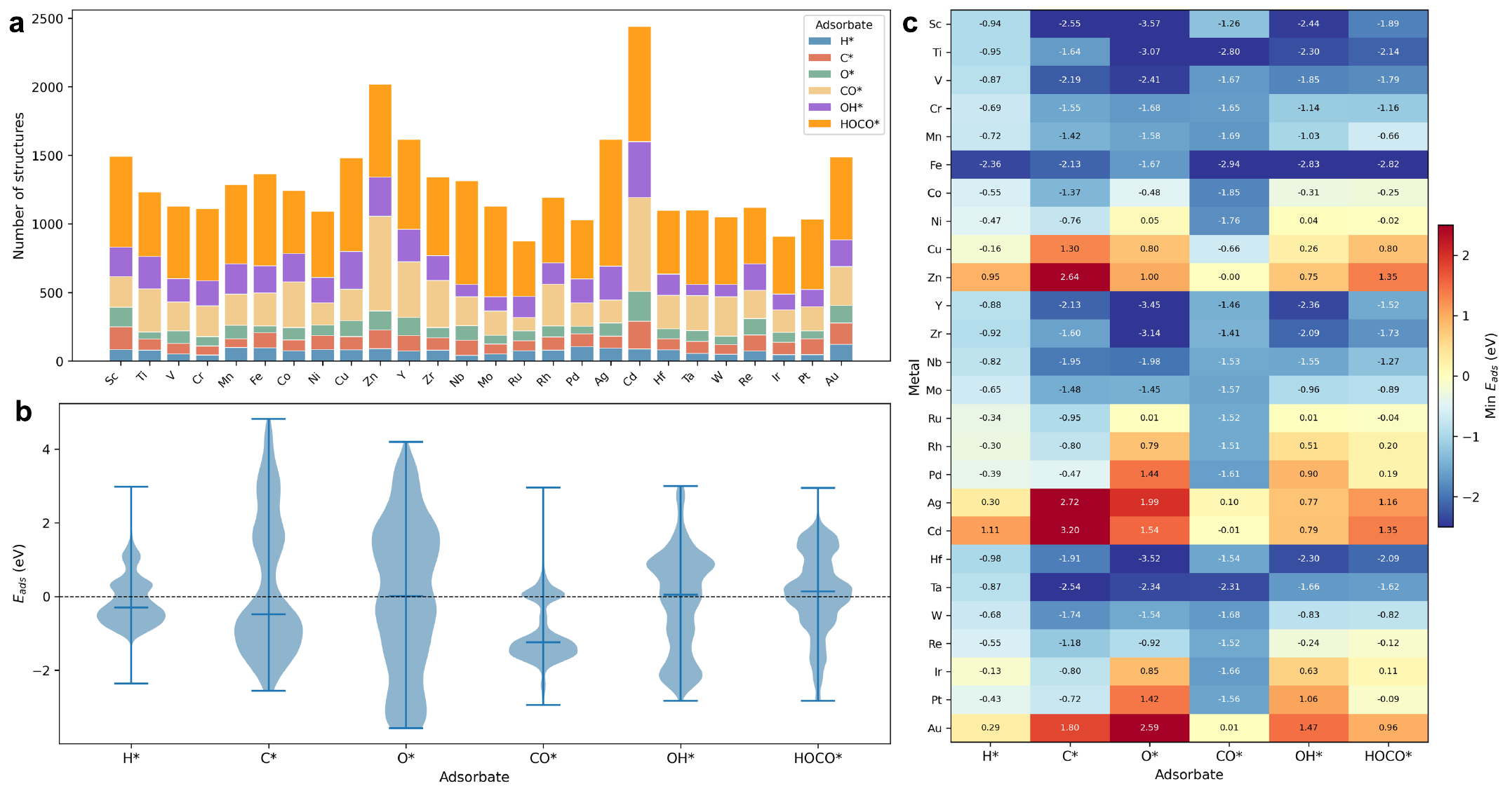}
    \caption{Dataset overview. (a) The number of structures for each metal and adsorbate in the dataset. (b) The adsorption energy distribution for each adsorbate. (c) The most stable adsorption energies for each metal and adsorbate.}
    \label{fig:data}
\end{figure}

Figure~\ref{fig:data}a summarizes the composition of AdsTraj across the 26 metals and six adsorbates, with per-combination counts listed in Table~\ref{tab:dataset}. The number of structures per metal ranges from 876 (Ru) to 2{,}507 (Cd). The dataset is weighted toward the configurationally richer adsorbates: HOCO* contributes 14{,}897 structures (44\% of the dataset) and CO* 7{,}044, whereas the monatomic adsorbates H*, C*, and O* contribute 1{,}984, 2{,}627, and 2{,}407 structures, respectively. This distribution reflects the larger number of distinct adsorption geometries available to polyatomic adsorbates, which differ not only in binding site but also in molecular orientation and internal conformation, and which therefore require denser sampling for the model to learn the directional information encoded in the adsorbate graph. The distribution of adsorption energies for each adsorbate across all metals is shown in Figure~\ref{fig:data}b. The distributions are broad, spanning 5--8~eV per adsorbate, which reflects the wide range of binding strengths accessible across the transition-metal series. O* exhibits the widest spread ($-3.6$ to $+4.2$~eV), consistent with its strong oxophilic interaction with early transition metals and its near-repulsive interaction with the coinage metals.\cite{Norskov2014,Nrskov2008} For CO*, 75\% of the sampled configurations have negative adsorption energies, the largest fraction among these adsorbates. The distributions of CO*, OH*, and HOCO* are visibly bimodal, separating strongly chemisorbed configurations from weakly bound or near-desorbed geometries that arise from the broad configurational sampling. The median adsorption energies of H*, C*, O*, CO*, OH*, and HOCO*, indicated by the horizontal lines in Figure~\ref{fig:data}b, are $-0.30$, $-0.48$, $0.01$, $-1.24$, $0.05$, and $0.14$~eV, respectively, with standard deviations of 0.66, 1.84, 1.98, 0.83, 1.36, and 1.09~eV. Figure~\ref{fig:data}c presents the most stable adsorption energy for each metal--adsorbate combination, which reveals a clear periodic trend. The early transition metals bind all six species strongly, most notably O* on Sc ($-3.57$~eV), Y ($-3.45$~eV), and Hf ($-3.52$~eV), whereas the
$d^{10}$ metals Zn, Ag, Cd, and Au bind weakly or not at all.\cite{Hammer1995} Binding strength weakens from left to right across each period as the $d$ band fills and its center shifts away from the Fermi level.\cite{Hammer2000,Norskov2014}

\subsection{Model Training}

\FloatBarrier
\begin{figure}[!ht]
    \centering
    \includegraphics[width=1.0\textwidth]{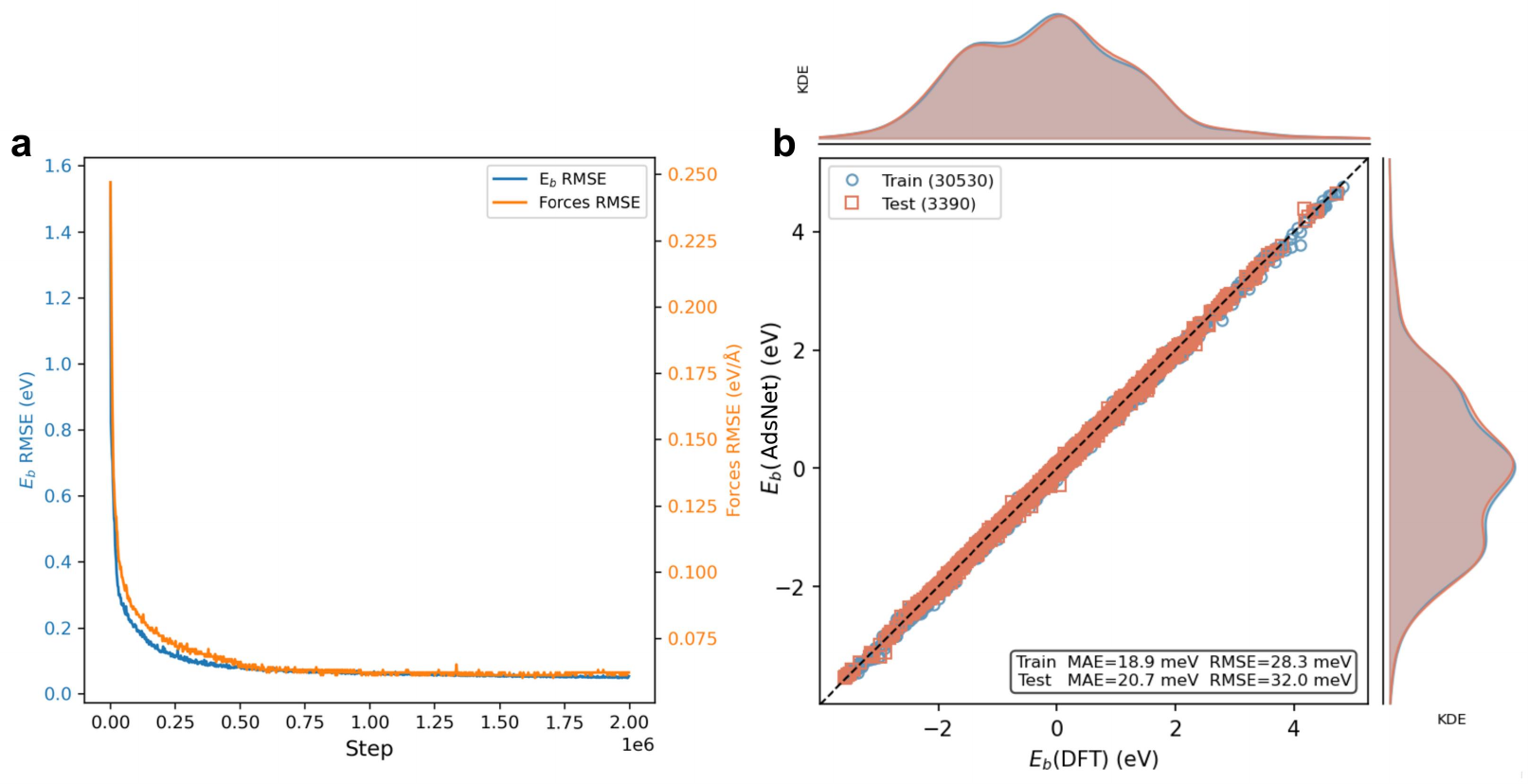}
    \caption{(a) Learning curves for the RMSE of adsorption energy and local atomic forces. (b) Parity plot comparing AdsNet-predicted adsorption energies against DFT-calculated values.}
    \label{fig:train}
\end{figure}

The model architecture employed in this work utilizes 3 interaction layers with 128 feature channels per node, a radial cutoff of 5.5~\AA{} expanded using sinc radial basis functions, and linear edge embeddings. The dataset is randomly split into a 90\% training partition and a 10\% test set, with 10\% of the training pool set aside for validation. We train AdsNet using the Adam optimizer with an initial learning rate of $10^{-4}$ and a batch size of 12. Training optimizes a multi-objective loss function combining the root-mean-square errors (RMSE) of the adsorption energy and atomic forces:
\begin{equation}
    \mathcal{L} = (1 - \lambda) \cdot \mathrm{RMSE}(E_{\text{ads}}) + \lambda \cdot \mathrm{RMSE}(\mathbf{F})
    \label{eqn:loss}
\end{equation}
where $\lambda = 0.9$ weights the force contribution to ensure accurate atomic forces for structural relaxation. Training proceeds for up to $2 \times 10^6$ steps, with the validation loss monitored every 2000 steps. Early stopping is triggered if the validation loss does not improve over 500 consecutive evaluation intervals.

The evolutions of the adsorption-energy and local-force RMSE during training are shown in Figure~\ref{fig:train}a. Both quantities drop by more than an order of magnitude within the first $2 \times 10^5$ steps and then decrease slowly, reaching 0.038~eV and 0.047~eV/\AA{} after $2 \times 10^6$ steps. Over the final $10^5$ steps the validation RMSE varies by less than 5~meV, confirming that training has converged. Although the loss in Eq.~\ref{eqn:loss} is dominated by the force term ($\lambda = 0.9$), the energy and force errors decrease on the same timescale, so the force weighting does not impede convergence of the adsorption energy. Figure~\ref{fig:train}b compares predicted and DFT adsorption energies for the 30{,}530 training (90\%) and 3{,}390 test (10\%) structures. The predictions follow the parity line across the full range of the dataset, from $-3.6$ to $+4.8$~eV, with RMSEs of 28.3 and 32.0~meV for the training and test sets, respectively. The test RMSE exceeds the training RMSE by only 3.7~meV, indicating that AdsNet generalizes to configurations outside the training set without appreciable overfitting. The kernel density estimates along both axes show that the predicted energy distribution reproduces the DFT distribution, including its two peaks at $0.05$ and $-1.5$~eV. The training and test distributions overlap closely, confirming that the random split is representative. In addition, the prediction accuracies by metal and adsorbate are also given in Tables~S1 and S2, which show that the test RMSE spans 19.8--41.1~meV across the 26 metals and 27.1--50.2~meV across the six adsorbates. More details on the training and test RMSEs for each metal and adsorbate are provided in the Supporting Information. The model's performance is consistent across different metals and adsorbates, demonstrating its generalizability for adsorption energy prediction on transition metal surfaces.

\subsection{Adsorption Energy contributions}

\begin{figure}[!ht]
    \centering
    \includegraphics[width=1.0\textwidth]{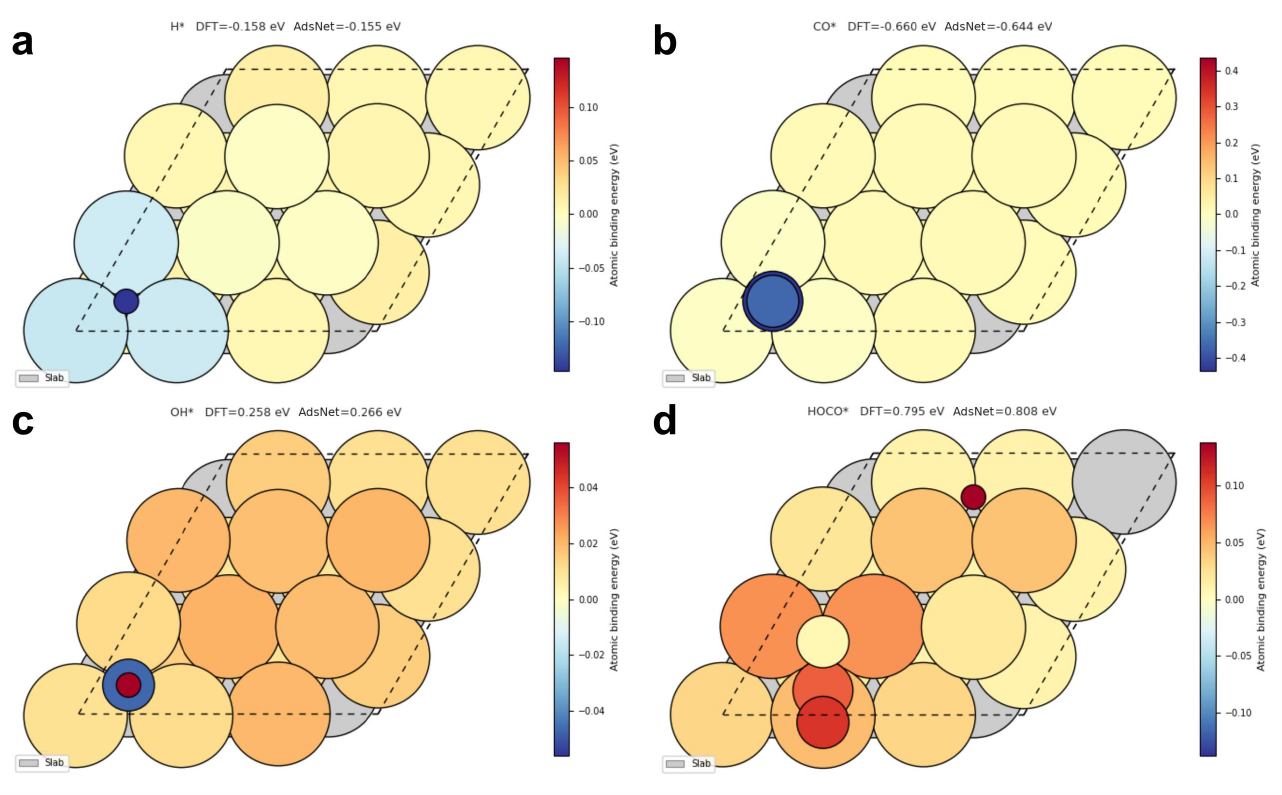}
    \caption{The predicted contributions plots of the local atoms within the local graph to total adsorption energies for (a) adsorbate H*, (b) adsorbate CO*, (c) adsorbate OH*, and (d) adsorbate HOCO*.}
    \label{fig:be_dist}
\end{figure}

By design, AdsNet naturally decomposes the predicted adsorption energy into individual atomic contributions within the local adsorbate graph, which is a capability that conventional MLIP methods cannot provide. Through the combination of learned scalar node features and the readout MLP, AdsNet resolves the total adsorption energy into site-specific contributions across the adsorbate and coordinating surface atoms. This provides insight into which atoms and interactions are most influential in determining the adsorption energetics. Intuitively, surface atoms directly coordinated to the adsorbate are expected to dominate the adsorption energetics relative to more distant neighbors; however, DFT total energies provide no direct physical quantity to measure this partitioning. By providing a tractable decomposition, AdsNet translates this physical intuition into a quantifiable scheme. This interpretability can help elucidate local structure–property relationships and guide rational catalyst design by pinpointing the key surface sites governing adsorption behavior. Figure \ref{fig:be_dist} shows the individual atomic contributions to the total adsorption energy across the H*, CO*, OH*, and HOCO* intermediates. The predicted adsorption energies closely match the DFT reference values across all cases, underscoring the high fidelity of AdsNet. Taking atomic hydrogen on Cu(111) as a representative single-atom adsorbate (Figure \ref{fig:be_dist}a), the partitioned atomic contributions within the local graph reveal that the H* adsorbate accounts for the dominant share of the adsorption energies ($-0.146$~eV). Coordinating surface copper atoms provide the next largest share of adsorption energy (approximately $-0.042$~eV), whereas more distant surface atoms exhibit negligible contributions (around $-0.008$~eV). This spatial decay confirms that adsorption energetics are governed primarily by the immediate local coordination environment. For diatomic adsorbates, we examine CO* and OH* on Cu(111) as representative benchmarks; their site-resolved atomic contributions to adsorption energy are displayed in Figures \ref{fig:be_dist}b and \ref{fig:be_dist}c, respectively.  For CO*, the C and O atoms account for $-0.436$~eV and $-0.367$~eV of the total adsorption energy, respectively. In the case of OH*, the adsorption energetic partition resolves into $-0.047$~eV for atom O and $0.056$~eV for atom H. Consistent with the trend observed for single atom H, the adsorbate atoms provide the dominant adsorption energetic contributions, followed by the coordinating surface atoms, while distant substrate atoms contribute marginally. To examine multi-atom intermediates, we consider the four-atom HOCO* intermediate on Cu(111) (Figure \ref{fig:be_dist}d). The decomposed atomic contributions across the adsorbate resolve into $0.138$, $0.109$, $0.089$, and $0.007$~eV for atoms H, O, C, and O, where the coordinating C and O atoms exhibit the strongest adsorption energetic contributions, while the distal atom H contributes comparatively less. These results demonstrate that AdsNet provides interpretable insights into the local bonding interactions that govern adsorption energetics, which can be valuable for understanding catalytic behavior and guiding the design of new catalysts.

\subsection{Structure Relaxation}

\begin{figure}[!ht]
    \centering
    \includegraphics[width=1.0\textwidth]{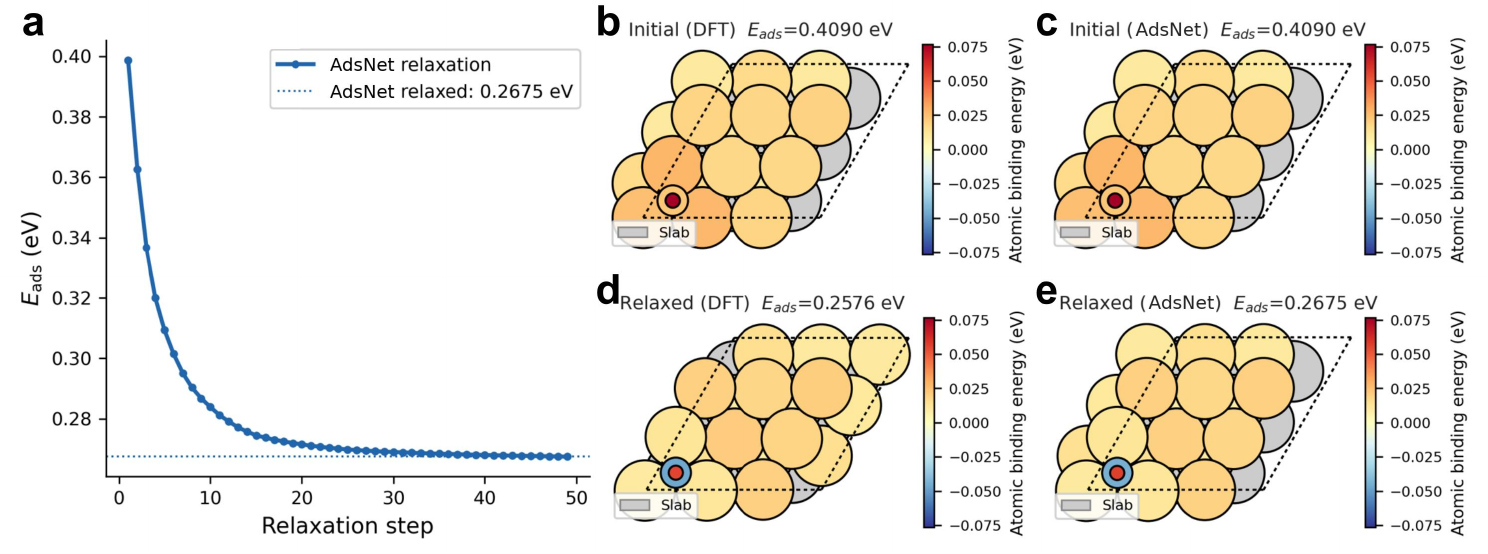}
    \caption{(a) AdsNet-accelerated local structure relaxation curve. (b) Initial and (d) final structures of DFT relaxation. (c) Initial and (e) final structures of AdsNet relaxation for OH* on Cu(111) surface.}
    \label{fig:relax}
\end{figure}

Local structure relaxation is crucial for accurate adsorption energetics, as the binding strength is governed directly by the adsorption geometry. AdsNet evaluates atomic forces via automatic differentiation, enabling rapid geometry optimization of adsorbate–surface interfaces. Leveraging these forces, local structure relaxation directly locates the minimum-energy adsorption configuration, bypassing expensive DFT iterations entirely. Figure \ref{fig:relax}a illustrates the process for AdsNet-accelerated local structure relaxation. Starting from an initial adsorption configuration, the model predicts atomic forces across the adsorbate and adjacent surface atoms, iteratively updating their coordinates until force convergence is reached. This localized scheme dramatically lowers the computational cost compared to conventional DFT structural relaxations.

The relaxation algorithm operates exclusively on local degrees of freedom defined within the adsorbate graph, optimizing only the coordinates of adsorbate atoms and their immediate surface neighbors. Box bounds are imposed on each coordinate to prevent unphysical displacements: the maximum displacement in the $xy$-plane is limited to $\Delta_{xy} = 0.5$~\AA{} per step, while the adsorbate $z$-coordinate is allowed a larger displacement of $\Delta_z = 1.0$~\AA{} to accommodate vertical adsorption height adjustments. The default optimization method is projected gradient descent (PGD) with an adaptive line search. At each iteration, the predicted forces (negative energy gradient) are used to propose trial steps at decreasing step sizes $\alpha \in \{0.5, 0.3, 0.1, 0.02, 0.01, 0.005, 0.001, 0.0005, 0.0001\}$. The trial position is projected onto the feasible box:
\begin{equation}
    \mathbf{r}_i^{\mathrm{trial}} = \mathrm{clip}\left(\mathbf{r}_i - \alpha \nabla_{\mathbf{r}_i} E_\mathrm{ads},\; \mathbf{r}_i^{\mathrm{lo}},\; \mathbf{r}_i^{\mathrm{hi}}\right)
    \label{eqn:pgd_step}
\end{equation}
and the first step size satisfying a sufficient decrease condition is accepted. If the line search fails to make progress, a coordinate-wise search is activated: individual degrees of freedom are perturbed along the direction of the largest gradient component with trial displacements of 0.05, 0.02, 0.01, 0.005, and 0.001~\AA{} to escape shallow local regions of the potential energy surface. Convergence is declared when either (i) the maximum force component falls below the force tolerance $\|\nabla E\|_\infty < 10^{-3}$~eV/\AA{}, (ii) the energy change remains below $10^{-5}$~eV for 30 consecutive iterations (energy plateau criterion), or (iii) the maximum number of model evaluations (80) is reached. Alternatively, L-BFGS-B can be used as the optimizer by interfacing with the same bounded objective function, which may provide faster convergence for smooth energy landscapes. In this work, we use the PGD method with a maximum of 80 model evaluations per relaxation.

Figures \ref{fig:relax}b--e demonstrate the workflow and performance of AdsNet-accelerated local structure relaxation using OH* on Cu(111) as a representative benchmark. The system is initialized from an unrelaxed configuration placed at an unoptimized, arbitrary surface site (Figures \ref{fig:relax}b and c). Driven by the analytic atomic forces evaluated via automatic differentiation through the localized adsorbate graph, AdsNet iteratively updates the Cartesian coordinates of the OH* species and its coordinating Cu surface atoms. Convergence is successfully reached after 49 iterations, yielding the optimized equilibrium geometry depicted in Figure \ref{fig:relax}e. The optimized configuration reaches a final adsorption energy of $0.268$~eV (Figure \ref{fig:relax}a), in excellent agreement with the fully DFT-relaxed reference geometry (Figure \ref{fig:relax}d) at $0.258$~eV. Notably, the local relaxation of AdsNet requires only a very small fraction of the computational cost compared to DFT relaxation, demonstrating both high efficiency and accuracy in predicting relaxed adsorption configurations.

\section{Conclusions}

We have presented AdsNet, an equivariant graph neural network designed for direct, rapid, and accurate prediction of adsorption energies with forces. By constructing representations over a local adsorbate graph rather than the full graph, AdsNet decouples the computational complexity from total system size, enabling rapid adsorption energy evaluation for large surface structures. Trained on 33,920 DFT-calculated configurations spanning 26 transition metals and 6 representative reaction intermediates (H*, C*, O*, CO*, OH*, and HOCO*), the AdsNet model delivers near-DFT accuracy, achieving a test-set adsorption energy RMSE of 0.032~eV, a force RMSE of 0.047~eV/\AA{}, and an $R^2$ of 0.999. By evaluating atomic forces via automatic differentiation, AdsNet drives efficient local structure relaxation using a projected gradient descent algorithm with adaptive line search at a negligible fraction of DFT computational cost. Furthermore, the architecture naturally decomposes the total adsorption energy into site-resolved atomic contributions—providing a quantifiable and interpretable lens into local bonding environments that cannot be extracted from standard DFT total energies or conventional MLIPs. Uniting the rapid interpretability of localized schemes with the high chemical precision, AdsNet delivers a scalable high-throughput paradigm for computational surface science. Future efforts will extend the framework to multimetallic alloy surfaces, high-coverage multi-adsorbate regimes, and automated transition-state search algorithms for surface science and catalyst discovery.

\section*{Declaration of Competing Interest}
The authors declare that they have no known competing financial interests or personal relationships that could have appeared to influence the work reported in this paper.

\section*{Data Availability}
The AdsTraj dataset is available at \url{https://huggingface.co/datasets/changzhiai/AdsTraj}. AdsNet model weights are available at \url{https://huggingface.co/changzhiai/adsnet-models}.

\section*{Code Availability}
The AdsNet source code is available at \url{https://github.com/changzhiai/AdsNet}. Documentation is available at \url{https://adsnet.readthedocs.io}. The version described in this paper is archived at \url{https://doi.org/10.5281/zenodo.24170856}.

\section*{Acknowledgment}
This work was supported by the U.S. Department of Energy, Office of Science, Office of Basic Energy Sciences, Chemical Sciences, Geosciences, and Biosciences Division, Catalysis Science Program to the SUNCAT Center for Interface Science and Catalysis. Computational resources were provided by the SLAC Shared Scientific Data Facility (S3DF) at SLAC National Accelerator Laboratory and the National Energy Research Scientific Computing Center (NERSC) at Lawrence Berkeley National Laboratory under the SUNCAT-FWP allocation supported by the Office of Science of the U.S. Department of Energy. Work also performed on the Carbon HPC cluster at the Center for Nanoscale Materials, a U.S. Department of Energy Office of Science User Facility, was supported by the U.S. DOE, Office of Basic Energy Sciences, under Contract No. DE-AC02-06CH11357.

\section*{Keywords}
Graph neural network; Machine learning potential; Adsorption energy; Catalysis; Metal surfaces; Adsorbates; AdsNet; AdsTraj; Equivariant message passing.

\end{document}


\maketitle
\newpage
\section*{Prediction Accuracy by Metal and Adsorbate}

Tables~\ref{tab:si_metal} and \ref{tab:si_ads} report the prediction errors of the trained AdsNet model resolved by surface metals and by adsorbates. Errors are computed from the same model and the same 90\%/10\% random split used for Figure~4b, and the last rows (All) in the Tables give the aggregate values. Across the 26 metals the test RMSE spans 19.8--41.1~meV, and across the six adsorbates 27.1--50.2~meV. The largest errors are obtained for Fe (41.1~meV) and Re (40.0~meV) among the metals, and for C* (50.2~meV) and O* (44.6~meV) among the adsorbates, which also span the widest adsorption-energy ranges in the dataset (7.4 and 7.8~eV, respectively).

\begin{table}[!ht]
\centering
\caption{Prediction errors of AdsNet by surface metal. $N$ is the number of structures for each metal; MAE and RMSE are given in meV.}
\label{tab:si_metal}
\small
\renewcommand{\arraystretch}{0.9}
\begin{tabular}{l r r r r r r}
\hline
 & \multicolumn{3}{c}{Train} & \multicolumn{3}{c}{Test} \\
Metal & $N$ & MAE & RMSE & $N$ & MAE & RMSE \\
\hline
Sc & 1355 & 17.4 & 23.8 & 137 & 19.4 & 26.5 \\
Ti & 1103 & 23.4 & 30.5 & 128 & 25.3 & 33.4 \\
V & 1004 & 19.0 & 31.0 & 124 & 19.1 & 30.0 \\
Cr & 1001 & 19.6 & 33.3 & 109 & 18.4 & 32.5 \\
Mn & 1162 & 15.4 & 24.0 & 123 & 18.1 & 29.5 \\
Fe & 1239 & 26.5 & 42.0 & 125 & 24.3 & 41.1 \\
Co & 1124 & 20.5 & 32.1 & 119 & 22.5 & 38.6 \\
Ni & 970 & 14.1 & 17.9 & 122 & 14.4 & 21.9 \\
Cu & 1330 & 12.1 & 17.6 & 151 & 13.3 & 19.8 \\
Zn & 1901 & 22.3 & 30.2 & 181 & 28.1 & 38.4 \\
Y & 1463 & 17.6 & 25.7 & 153 & 20.9 & 37.0 \\
Zr & 1210 & 17.6 & 24.2 & 130 & 19.8 & 33.2 \\
Nb & 1180 & 15.7 & 22.2 & 134 & 18.6 & 24.4 \\
Mo & 1018 & 16.8 & 25.2 & 110 & 16.4 & 26.4 \\
Ru & 781 & 16.3 & 23.7 & 95 & 18.0 & 25.3 \\
Rh & 1066 & 16.3 & 22.6 & 126 & 16.7 & 23.5 \\
Pd & 934 & 21.8 & 35.9 & 93 & 19.4 & 32.6 \\
Ag & 1445 & 20.7 & 30.5 & 170 & 24.5 & 38.9 \\
Cd & 2258 & 22.2 & 34.4 & 249 & 26.1 & 38.2 \\
Hf & 984 & 23.0 & 28.5 & 113 & 22.5 & 26.7 \\
Ta & 998 & 16.4 & 23.5 & 102 & 16.0 & 22.3 \\
W & 937 & 16.5 & 23.1 & 112 & 19.8 & 30.1 \\
Re & 1002 & 19.4 & 27.2 & 118 & 21.4 & 40.0 \\
Ir & 806 & 16.7 & 24.0 & 102 & 20.9 & 31.1 \\
Pt & 930 & 19.3 & 28.1 & 105 & 21.5 & 33.3 \\
Au & 1329 & 18.4 & 31.0 & 159 & 21.1 & 29.8 \\
\hline
All & 30530 & 18.9 & 28.3 & 3390 & 20.7 & 32.0 \\
\hline
\end{tabular}
\renewcommand{\arraystretch}{1.0}
\end{table}

\begin{table}[!ht]
\centering
\caption{Prediction errors of AdsNet by adsorbate. $N$ is the number of
structures for each adsorbate; MAE and RMSE are given in meV.}
\label{tab:si_ads}
\small
\renewcommand{\arraystretch}{0.9}
\begin{tabular}{l r r r r r r}
\hline
 & \multicolumn{3}{c}{Train} & \multicolumn{3}{c}{Test} \\
Adsorbate & $N$ & MAE & RMSE & $N$ & MAE & RMSE \\
\hline
H* & 1798 & 19.8 & 31.1 & 186 & 22.2 & 36.6 \\
C* & 2368 & 25.0 & 35.1 & 259 & 31.2 & 50.2 \\
O* & 2174 & 27.1 & 39.5 & 233 & 30.1 & 44.6 \\
CO* & 6383 & 18.5 & 27.8 & 661 & 19.0 & 29.3 \\
OH* & 4435 & 17.5 & 27.0 & 526 & 18.0 & 27.1 \\
HOCO* & 13372 & 17.0 & 24.9 & 1525 & 18.9 & 27.6 \\
\hline
All & 30530 & 18.9 & 28.3 & 3390 & 20.7 & 32.0 \\
\hline
\end{tabular}
\renewcommand{\arraystretch}{1.0}
\end{table}